\documentclass[preprint,12pt]{elsarticle}
\usepackage{graphicx,amsmath}
\journal{PHYSICA A}
\begin{document}

\title{A thermodynamic counterpart of the Axelrod model of social influence: The one-dimensional case}

\author[ivic,cpc]{Y. Gandica}
\ead{ygandica@gmail.com}
\author[ivic]{E. Medina}
\author[ivic]{I. Bonalde}

\address[ivic]{Centro de F\'{\i}sica, Instituto Venezolano de
Investigaciones Cient\'{\i}ficas, Apartado 20632, Caracas
1020-A, Venezuela}
\cortext[cor1]{Corresponding author}
\address[cpc]{Center for Computational Physics, Departamento de F\'{\i}sica, Universidade de Coimbra.
3004-516 Coimbra. Portugal}

\date{\today}

\begin{abstract}
We propose a thermodynamic version of the Axelrod model of
social influence. In one-dimensional (1D) lattices, the
thermodynamic model becomes a coupled Potts model with a bonding
interaction that increases with the site matching traits. We analytically
calculate thermodynamic
and critical properties for a 1D system and show
that an order-disorder phase transition only occurs at $T=0$
independent of the number of cultural traits $q$ and features
$F$. The 1D thermodynamic Axelrod model
belongs to the same universality class of the Ising and Potts
models, notwithstanding the increase of the internal dimension
of the local degree of freedom and the state-dependent bonding
interaction. We suggest a unifying proposal to compare exponents
across different discrete 1D models. The comparison with our Hamiltonian
description reveals that in the thermodynamic limit the original
out-of-equilibrium 1D Axelrod model with noise behaves like an ordinary
thermodynamic 1D interacting particle system.
\end{abstract}

%\pacs{89.20.-a, 89.75.Hc}

\maketitle

\section{Introduction}

The Axelrod model \cite{axelrod} was proposed originally to study dissemination of cultures among
interacting individuals or agents. Although the model is too simple to simulate social dynamics,
the mechanisms used in the model have been
recognized by social scientists as a global self-reinforcing social dynamic \cite{castellano2}. It is a
fact that the more culturally similar the people, the greater the chance of interaction between them, and that interaction increases their similarity \cite{barrat}. These are the premises of the model.

More explicitly, the Axelrod model considers that an agent located at the
$i^{th}$ site of a lattice is defined by a set of $F$ cultural
features (e.g., religion, sports, politics, etc.) represented
by a vector
$\sigma_{i}=(\sigma_{i1},\sigma_{i2},...,\sigma_{iF})$. Each
feature $\sigma_{ik}$ can take integer values in the interval
$[1,q]$, where $q$ defines the cultural traits allowed per
feature and measures the cultural variability in the system.
There are $q^F$ possible cultural states. The model's dynamics is as follows: (1) Choose randomly two
nearest neighbor agents $i$ and $j$, then (2) calculate the
number of shared features between the agents
$\ell_{ij}=\sum_k^F \delta_{\sigma_{ik},\sigma_{jk}}$. If
$0<\ell_{ij}< F$, then (3) pick up randomly a feature $k$ such
that $\sigma_{ik}\neq \sigma_{jk}$ and with probability
$\ell_{ij}/F$ set $\sigma_{ik} = \sigma_{jk}$. These time steps
are iterated and the dynamics stops when a frozen state is
reached; i.e., either $\ell_{ij}=0$ or $\ell_{ij}=F, \, \forall
i,j$. A cluster is a set of connected agents with the same
state. Monocultural or ordered phases are composed of a cluster
of the size of the system where $\ell_{ij}=F, \forall i,j$.
Multicultural or disordered phases consist of two or more
clusters.

One of the main features of this model is a change of behavior
at a value $q_c$ from a monocultural state, where all agents
share the same cultural features, to a multicultural state,
where individuals mostly have their own features \cite{castellano1}. This
change can be characterized by an order parameter $\phi$ that
is usually defined as the average size of the largest cultural
cluster $C_{\textrm{max}}$ normalized by the total number of
agents $N$ in the system; $\phi=C_{\textrm{max}}/N$. In the
monocultural (ordered) state $\phi\rightarrow1$ and in the
multicultural (disordered) state $\phi\rightarrow0$.

The insertion of additional ingredients in the model, like an external field or mass media, yields interesting nontrivial consequences in the system \cite{gonzalez1,gonzalez2,yg}. The main limitation of the model seems to be that the system always converges to absorbent states, a situation that clearly does not occur in society.
Some variants of the model relax this tendency by introducing noise into the system \cite{economia,klemm2,toral}. If the noise rate is small, the system reaches only monocultural states. However, if the noise rate is above a size-dependent critical value, a polarized state is sustained \cite{economia,klemm2,toral}. Klemm \textit{et al}. \cite{economia,klemm2}
associated the monocultural (multicultural) states with stable (unstable) equilibria.

Until now, in the sociophysics field the global dynamics of social systems have been usually studied by postulating a series of rules that at the end lead to out-of-equilibrium behaviors, such as
absorbent states. This approach often uses statistical mechanics concepts -temperature, critical phase transition, applied magnetic field, among others- without formal definitions. Langevin-type approaches have been proposed to study the collective phenomena of the social systems in terms of their microscopic constituents and their interactions \cite{hammal}. Little attention has been paid to
this approach in which the system can be modeled in a Hamiltonian formulation, whereupon
equilibrium and nonequilibrium behaviors can be explored \cite{report,Nonequilibrium}. Such a Hamiltonian description also allows for the understanding of the meaning of social variables in the context of statistical mechanics.

Here, we develop a Hamiltonian version of the Axelrod model of social influence. Our Hamiltonian captures the local interactions of the original model. With the aim of finding a
possible thermodynamic role of the parameters $F$, $q$, and $q_c$, our model, henceforth called \textit{thermodynamic Axelrod}, uses the number of shared features $\ell_{ij}$ of the Axelrod model to construct a new Hamiltonian distinguishable from the 1D $F$-parallel-layer Potts models in that the interaction strength between agents increases with $\ell_{ij}$. This feature of the interaction precipitates ordering preempting fluctuations. In the thermodynamic Axelrod model $F$ is related to the coupling energy of the system and $q$ has the same meaning as in the Potts model.

Although it is usually argued that the Axelrod model is an out-of-equilibrium model, this fact, taken as obvious, has never been demonstrated in the literature. In Sect.~II we
demonstrate that the standard Axelrod model does not satisfy the detailed balance condition.
In Sect.~III we analytically calculate the main thermodynamic functions for our model. For the critical
behavior analysis (Sect.~IV-b) we make an unifying proposal to compare exponents across different 1D discrete models, since current definitions depend on model details. In Sect.~V we state the consequences of our study over the transitions driven by
noise in the original Axelrod model. We discuss the implications of a thermodynamic society in
Sect.~VI and in Sect.~VII we present our conclusions.

\section{Axelrod model: Out of equilibrium}
Before we introduce the thermodynamic version of the Axelrod
model, here we demonstrate that the original version does not
satisfy equilibrium conditions by showing that
detailed balance is violated.

Let a link between two sites $i$ and $j$ be of type $n$ if they
share $n$ components ($\ell_{ij}=n$), $P_n$ be the probability
that the system is in a state with links of type $n$, and
$W_{nm}$ be the transition probability per unit time from a
state with type-$n$ to one with type-$m$ links. $W_{nm}$ being
time independent. Since in the dynamics of the Axelrod model a
feature $k$ is changed with probability $\ell_{ij}/F$ to make
two sites have one more component in common
($\sigma_{ik}=\sigma_{jk}$), we have

\begin{equation}
W_{nm}= \left \{ \begin{array}{r@{\qquad}c@{\qquad}l}
\frac{n}{F} & \mbox{for} & m=n+1, \\ 0 & & \mbox{otherwise.}
\end{array} \right.
\label{prob}
\end{equation}

\noindent The \textit{detailed balance} relation implies that \cite{diu}

\begin{equation}
W_{nm}P_m = W_{mn}P_n  \qquad \forall \quad n,m \,. \label{dbe}
\end{equation}

\noindent In the Axelrod model Eq.~(\ref{dbe}) cannot be
satisfied since one side or the other is always zero according
to Eq.~(\ref{prob}). This feature of the model introduces some
very strong constraints into both the evolution and the
time-independent states of interactions that emphasize
nonequilibrium; i.e. i) completely different individuals do not
interact, ii) individuals conform once they have modified
their cultural profile, and iii) individuals that are alike,
that interact, increase their similarity at interaction. These
rules yield absorbing states, the most salient nonequlibrium
feature of the Axelrod model.

The thermodynamic model we propose relaxes all of the previous
constraints, while preserving similarity, increasing
interactions, and introducing fluctuations controlled by a
temperature parameter. These conditions permit arriving at a
dynamical equilibrium state in the regular sense of statistical
mechanics. On the other hand, we are interested in evaluating
whether behaviors reported in nonequilibrium network models
survive in a thermodynamic driven scenario.

\section{Thermodynamic Axelrod model}

To reproduce the interaction rule of the Axelrod model, in which
the interaction probability is proportional to the number of
shared features, the Hamiltonian is defined as
\begin{equation}\mathcal{H} =
-\sum_{k=1}^F \sum_{ij}^N \left(
J_{ij}\,\delta(\sigma_{ik},\sigma_{jk}) + \frac{\mu
H}{2}\left[\delta(\sigma_{ik},H_{k}) +
\delta(\sigma_{jk},H_{k})\right] \right) \,, \label{hamil}
\end{equation}

\noindent with the interaction factor

\begin{equation}
J_{ij}=\sum_{n=1}^F J \delta(\sigma_{in},\sigma_{jn}) \,.
\label{jint}
\end{equation}

\noindent The delta function captures the local interactions of the original model, inasmuch as the interaction strength between agents increases with the number of shared features $\ell_{ij}$. In this way our Hamiltonian system takes into account both the tendency of individuals to become more similar when they interact, namely social influence (like the voter model), and the greater tendency to interact with individuals which are more similar, namely homophyly (specific of the
Axelrod model). $H_k$ works as an applied magnetic field that (a)
can point in one of the Potts-model-like directions $k$, (b) can
take values $1,...,q$, and (c) has an energy weight proportional to
the magnitude $H$. $\mu$ is the magnetic-like moment per agent.
$\sigma_{ik} = 1, \ldots, q$ specifies each
of the $F$ variables $(\sigma_{i1}, \sigma_{i2},\ldots,
\sigma_{iF})$ of the agent $\sigma_i$ at the $i$th lattice
site. $N$ is the size of the system. The second term in the
Hamiltonian is symmetrized for convenience.

The Hamiltonian in Eq.~(\ref{hamil}) is evidently inspired on the Potts model, with
the significant distinction that the \emph{interaction factor
$J_{ij}$ always depends on the global state of the F-vector}
and not on the state of the particular Potts variable. This is a
somewhat rare type of Hamiltonian interaction, in a sense
similar to a nonlinear sigma model where a vector interaction
occurs subject to normalization of the interacting vectors
\cite{kardar}. Our model on a 1D lattice is like an $F$-coupled-layer
Potts model. It is then a quasi 1D system, thus some signatures of the two-dimensional (2D) Potts model are expected as crossovers.

We consider a 1D chain of $N$ sites occupied by $q^F$-valued
agents and use the transfer matrix method to compute the
relevant physical properties. For periodic boundary conditions
$\sigma_{(N+1)k}=\sigma_{1k}$, the partition function
corresponding to the above Hamiltonian can be expressed as

\begin{eqnarray}
Z & = & \sum_{\sigma_1}\sum_{\sigma_2} \cdots \sum_{\sigma_N}
\prod_{i=1}^N \exp \Bigg[ \sum_{k=1}^F \bigg(\beta
J_{i(i+1)}\delta(\sigma_{ik},\sigma_{(i+1)k}) \nonumber\\
&  & + \frac{\beta \mu H}{2}\Big[\delta(\sigma_{ik},H_{k}) +
\delta(\sigma_{(i+1)k},H_{k}) \Big] \bigg) \Bigg] \nonumber\\
& = & Tr[W^N] \,. \label{partition}
\end{eqnarray}

\noindent Here, $\beta=1/k_BT$ and we introduced a $q^F \times
q^F$ transfer matrix $W$ with elements

\begin{eqnarray}
\langle\sigma_i|W|\sigma_{i+1}\rangle & = &\exp \Bigg[
\sum_{k=1}^F \bigg(\beta
J_{i(i+1)}\delta(\sigma_{ik},\sigma_{(i+1)k}) \nonumber\\
&  &  + \frac{\beta \mu H}{2}\Big[\delta(\sigma_{ik},H_{k}) +
\delta(\sigma_{(i+1)k},H_{k}) \Big] \bigg) \Bigg] \,.
\label{elem}
\end{eqnarray}

\noindent The eigenvalues $\lambda_j$ of the transfer matrix
are determined from the solution of the secular equation ${\rm Det}|W
-\lambda E|=0$. Then, the partition function can be written as

\begin{equation}
Z=\lambda_1^N + \lambda_2^N + \cdots \lambda_{q^F}^N =
\lambda_{max}^N \left(1 + \frac{\lambda_1^N}{\lambda_{max}^N} +
\cdots + \frac{\lambda_{q^F}^N}{\lambda_{max}^N} \right) \,,
\label{partlambda}
\end{equation}

\noindent where $\lambda_{max}$ is the largest eigenvalue. In
the thermodynamic limit ($N\rightarrow \infty$),
$Z\cong\lambda_{max}^N$. Then, a standard procedure
\cite{pathria} can be followed to obtain thermodynamic and
critical properties.

\section{Case $F=2$, $q=2$}

Here, we present explicitly the simplest nontrivial case F = 2,
q = 2, as an illustration. Analytical
calculations were performed for higher values of F and q but full expressions are
too lengthy. The
transfer matrix takes the form

\begin{equation} W= \left( \begin{array}{cccc} e^{2\beta \mu H + 4 \beta J}
& e^{\frac{3}{2} \beta \mu H+\beta J} & e^{\frac{3}{2} \beta
\mu H+\beta
J} & e^{\beta \mu H} \\
e^{\frac{3}{2} \beta \mu H+\beta J} & e^{\beta \mu H + 4 \beta
J} &
e^{\beta \mu H} & e^{\frac{1}{2} \beta \mu H+\beta J} \\
e^{\frac{3}{2} \beta \mu H+\beta J} & e^{\beta \mu H} &
e^{\beta \mu H + 4 \beta J} & e^{\frac{1}{2} \beta \mu H+\beta J}  \\
e^{\beta \mu H} & e^{\frac{1}{2} \beta \mu H+\beta J} &
e^{\frac{1}{2} \beta \mu H+\beta J} & e^{4 \beta J}
\end{array} \right) \,. \label{matrix22}
\end{equation}

\noindent The largest eigenvalue for this matrix is

\begin{eqnarray}
\lambda_{max} & = & \frac{1}{3} \left (e^{4\beta J + \beta \mu
H} + e^{4\beta J + 2\beta \mu H} + e^{4\beta J} +e^{\beta \mu
H} \right) - \frac{2^{1/3}a}{3 \left[
b+\sqrt{4a^3+b^2}\right]^{1/3} } \nonumber \\
&  & + \frac{1}{3\;2^{1/3}} \left[
b+\sqrt{4a^3+b^2}\right]^{1/3} \,, \label{largeigen}
\end{eqnarray}

\noindent where

\begin{eqnarray*}
a & = & -6e^{2\beta J + \beta \mu H} + e^{4\beta J + \beta \mu
H} + e^{8\beta J + \beta \mu H} - 2e^{4\beta J + 2\beta \mu H}
- 6e^{2\beta J + 3\beta \mu H}
\\ & & + e^{4\beta J + 3\beta \mu H} +
e^{8\beta J + 3\beta \mu H} - e^{8\beta J + 4\beta \mu H} -
e^{8\beta J} - 4e^{2\beta \mu H}
\end{eqnarray*}

\begin{eqnarray*}
b & = & 18e^{6\beta J + \beta \mu H} - 3e^{8\beta J + \beta \mu
H} - 3e^{12\beta J + \beta \mu H} + 18e^{2\beta J + 2\beta \mu
H} + 6e^{4\beta J + 2\beta \mu H} \nonumber\\ &  & +
18e^{6\beta J + 2\beta \mu H} - 6e^{8\beta J + 2\beta \mu H} -
6e^{12\beta J + 2\beta \mu H} + 108e^{2\beta J + 3\beta \mu H}
- 12e^{4\beta J +
3\beta \mu H} \nonumber\\
 &  & - 72e^{6\beta J + 3\beta \mu H} + 18e^{8\beta J +
3\beta \mu H} + 14e^{12\beta J + 3\beta \mu H} + 18e^{2\beta J
+
4\beta \mu H} + 6e^{4\beta J + 4\beta \mu H} \nonumber \\
&  & + 18e^{6\beta J + 4\beta \mu H} - 6e^{8\beta J + 4\beta
\mu H} - 6e^{12\beta J + 4\beta \mu H} + 18e^{6\beta J + 5\beta
\mu H} -
3e^{8\beta J + 5\beta \mu H} \nonumber \\
&  & - 3e^{12\beta J + 5\beta \mu H} + 2e^{12\beta J + 6\beta
\mu H} + 2e^{12\beta J} - 16e^{3\beta \mu H} \,.
\end{eqnarray*}

The free energy, magnetization, magnetic susceptibility per particle, and
specific heat are given in terms of $\lambda_{max}$:

\begin{equation}
\mathcal{F}=-k_B T \ln \lambda_{max} \,, \label{freeenergy}
\end{equation}
\begin{equation}
M=-\frac{\partial \mathcal{F}}{\partial H}=\frac{k_BT
}{\lambda_{max}}\frac{\partial\lambda_{max}}{\partial H} \,,
\label{magn}
\end{equation}
\begin{equation}
\chi=\frac{\partial M}{\partial H}=\frac{\partial}{\partial H}
\left (
\frac{k_BT}{\lambda_{max}}\frac{\partial\lambda_{max}}{\partial
H} \right ) \,, \label{suscep}
\end{equation}
\begin{equation}
C=-T\frac{\partial^2 \mathcal{F}}{\partial
T^2}=2\frac{k_BT}{\lambda_{max}}\frac{\partial\lambda_{max}}{\partial
T}+ k_BT^2 \frac{\partial}{T} \left
(\frac{1}{\lambda_{max}}\frac{\partial \lambda_{max}}{\partial
T} \right ) \,. \label{spheat}
\end{equation}
\begin{figure}
\unitlength1in
\begin{minipage}[t]{3in}
\scalebox{0.25}{\includegraphics{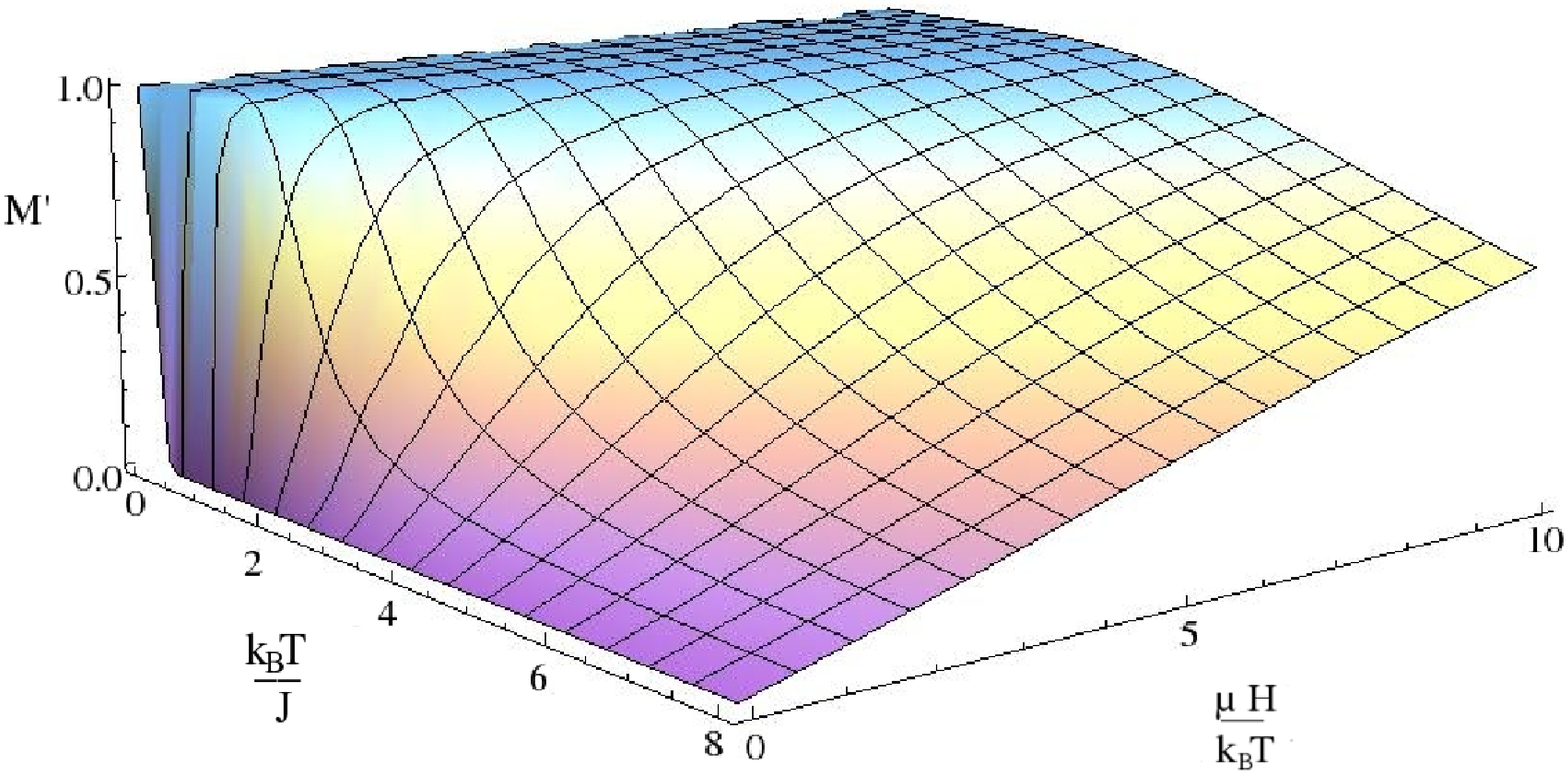}}\par
\caption{Magnetization of the 1D thermodynamic Axelrod model
for the case $F=2$ and $q=2$.} \label{f1}
\end{minipage}
\hfill
\begin{minipage}[t]{3in}
\scalebox{0.25}{\includegraphics{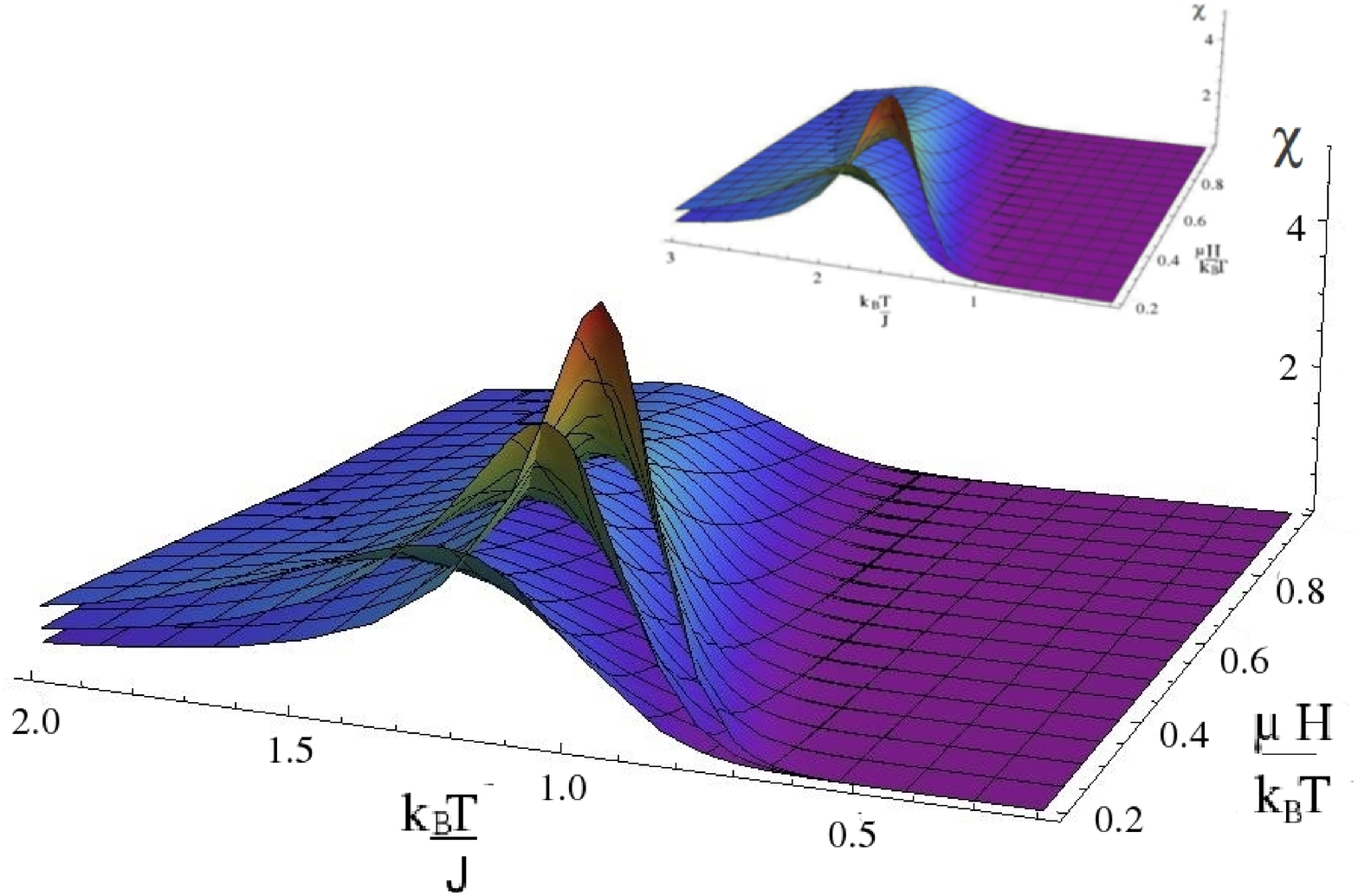}}\par
\caption{Susceptibility of the 1D thermodynamic Axelrod model
for the case $F=2$ with $q=2,3$ and $4$. In the inset we show the case $F=3$ with $q=2$ and $3$.} \label{f2}
\end{minipage}
\vfill
\begin{minipage}[t]{3in}
\scalebox{0.29}{\includegraphics{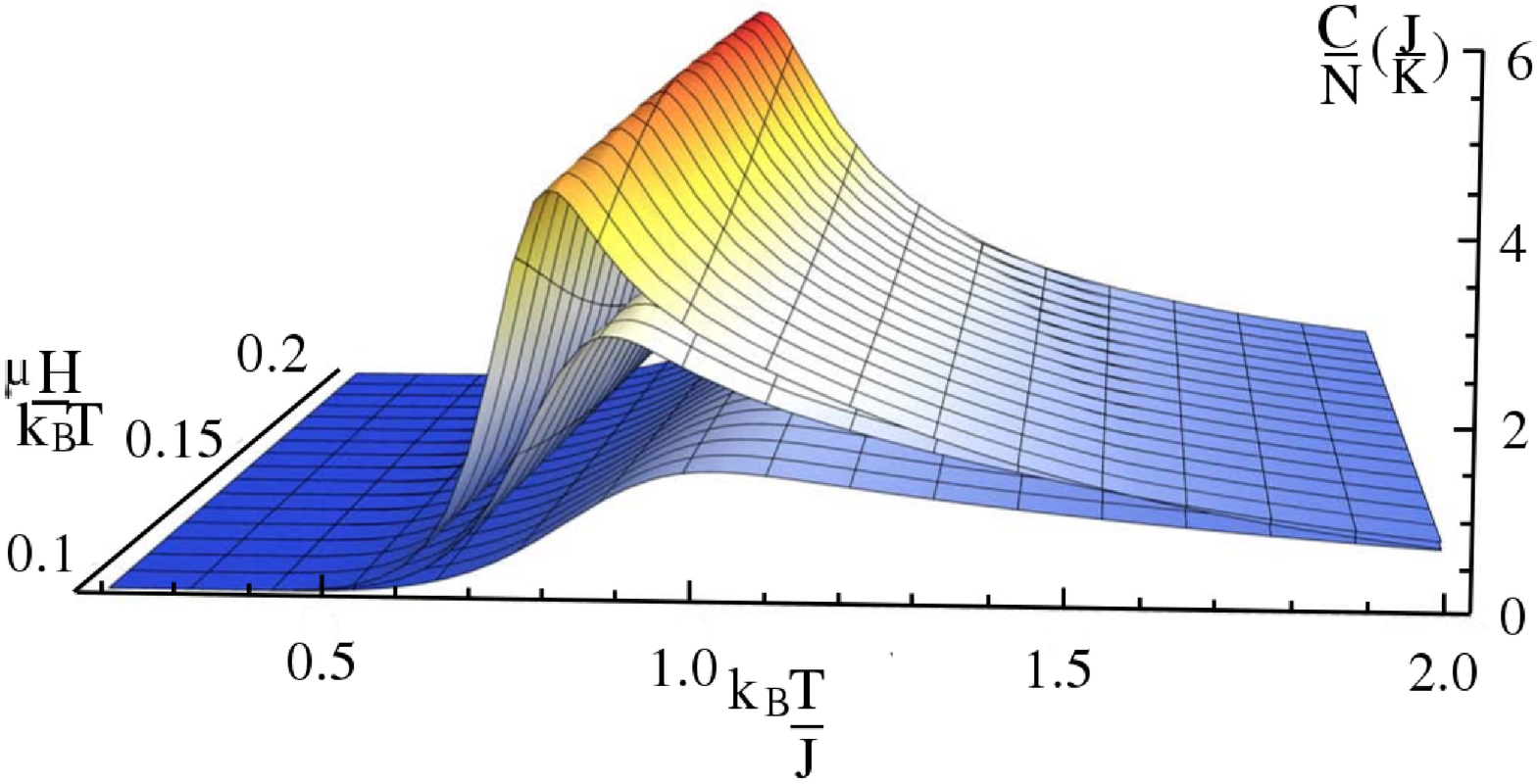}}\par
\caption{Specific heat of the 1D thermodynamic Axelrod model for
the case $F=2$ and $q=2,3$ and $4$.} \label{cef2}
\end{minipage}
\hfill
\begin{minipage}[t]{3in}
\scalebox{0.33}{\includegraphics{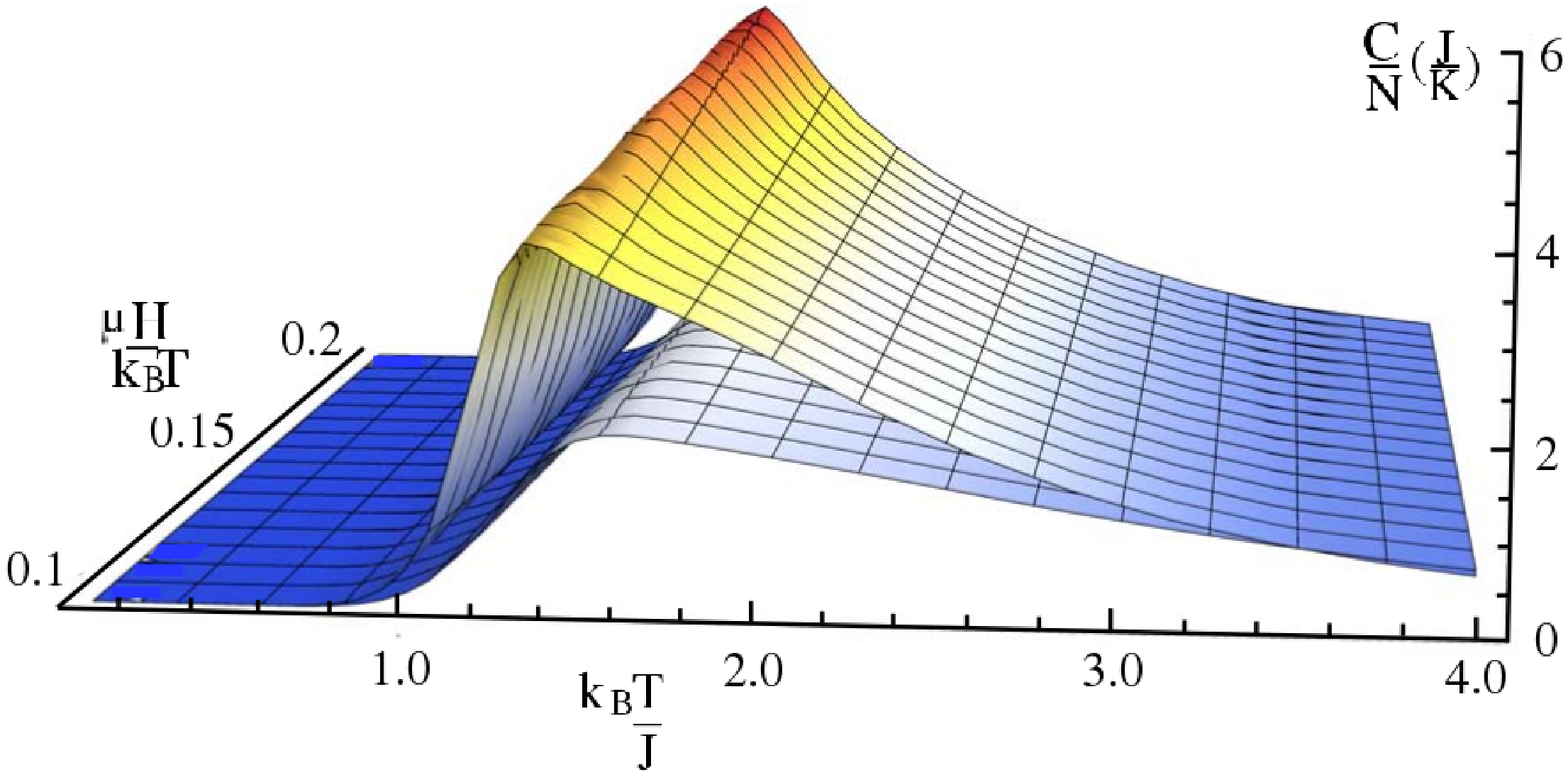}}\par
\caption{Specific heat of the 1D thermodynamic Axelrod model for
the case $F=3$ and $q=2$ and $q=3$.} \label{cef3}
\end{minipage}
\end{figure}
\noindent Figures~\ref{f1}-\ref{cef3} show plots of the
magnetization, magnetic susceptibility, and specific heat in
terms of $k_B T/J$ and $\mu H/k_B T$. The magnetization (Fig.~\ref{f1})
goes to 0 as $H\rightarrow0$
at any finite $T$. At $T=0$, the magnetization saturates to its
maximum value for any $H$. This implies a spontaneous
transition to an ordered state only at $T=0$. When $T$ is
finite, the magnetization saturates to its maximum only at
large $H$. Thus, the changes in the internal space dimensions of the model do not affect the
scenario expected for one dimension. The increased fluctuations derived from the
greater dimensions of the internal vector space destroy order except at $T=0$,
as would be expected.

The susceptibility diverges as $T\rightarrow0$ and $H\rightarrow0$. As we will derive analytically in the following section, the divergence is independent of the values $F$ and $q$ and corresponds to the Potts class.  On the other hand, the nonuniversal prefactors become larger as the value of $q$ increases, as can be seen in Fig.\ref{f2}. The farther from the singularity in the $T$ direction, the faster high-$q$ susceptibilities die out. The opposite is true in the $H$ direction. This is the scenario
in both $F=2$ and $F=3$.

As expected, the temperature of the maximum value of the specific heat $C$ (Schottky anomaly)
increases for larger $F$ values (Fig.~\ref{cef2} and Fig.~\ref{cef3}). As a
result of the coupling, the ordered system is more
robust and requires more energy to be destroyed. For a fixed $F$ value, $C$ depends only on the external field at different values of $q$. Since the specific heat is proportional to the amount of energy per agent the system can absorb, as $q$ increases the number of accessible states is greater
and, therefore, a higher external field is required to orient the agents in the
same direction. The dependence of $C$ with temperature is through the
the gap of the system (difference between the ground state and the first excited
state), which does not vary with $q$ (see Eq.~(\ref{expt}) below). All
these properties behave qualitatively as they do in the Ising and Potts models
on 1D lattices.

The specific heat can be regarded as the resistance of the society to increase the fluctuation average in posture. For lower values of $q$ it is
easier to change the average size of fluctuations, because there are
less options for disagreement in the system. The gap, which increases with $F$, is
the energy necessary for the system to break similarity bonds between individuals
in the ground state.

The susceptibility is the magnetic response of the system
to external field. In the thermodynamic society, it represents the relative ease of
social alignment to the mass media. As thermal fluctuations decrease,
a weaker mass media makes for a larger effect. This can be seen from the fact that as $q$ increases
the system magnetizes more easily at smaller fields.

In terms of the thermal society of agents, the above results
imply that, as a consequence of the periodic boundary conditions,
the fluctuating postures will always outweigh the benefits of agreement so that
spontaneous cultural uniformity does not occur. Uniformity can occur, however, for any finite mass media.

\subsection{Spatial correlations}

We now calculate the two-point correlation function
\begin{equation}
G(i,i+j)=\langle \sigma_i\sigma_{i+j}\rangle
-\langle\sigma_i\rangle\langle\sigma_j\rangle \label{corrf}
\end{equation}
using the transfer matrix method \cite{goldenfeld}. The first term is given by
\begin{equation}
\langle\sigma_i\sigma_{i+j}\rangle=\frac{1}{Z}{\rm tr}[AW^jAW^{N-j}]
\label{sigmatwo} \,,
\end{equation}
where
$A=\sum_{\sigma_i}|\sigma_i\rangle\sigma_i\langle\sigma_i|$. In
the original space for $F=2$ and $q=2$, the state matrix
\begin{equation}
A=\left(
    \begin{array}{cccc}
      1 & 0 & 0 & 0 \\
      0 & 2 & 0 & 0 \\
      0 & 0 & 2 & 0 \\
      0 & 0 & 0 & 4 \\
    \end{array}
  \right) \,.
\end{equation}
Following a standard procedure, we evaluate
Eq.~(\ref{sigmatwo}) in a basis where $W$ is diagonal. For
$H=0$, the unitary matrix that diagonalizes $W$ is
\begin{equation}
P=\left(
    \begin{array}{cccc}
      -1 & 0 & 1 & 1 \\
      0 & -1 & -1 & 1 \\
      0 & 1 & -1 & 1 \\
      1 & 0 & 1 & 1 \\
    \end{array}
  \right) \,.
\end{equation}
The evaluation yields
\begin{eqnarray}
\langle\sigma_i\sigma_{i+j}\rangle & = & \frac{1}{Z}{\rm tr} \Biggm[ \left(
    \begin{array}{cccc}
      \frac{5}{2} & 0 & \frac{3}{2} & \frac{3}{2} \\
      0 & 2 & 0 & 0 \\
      \frac{3}{4} & 0 & \frac{9}{4} & \frac{1}{4} \\
      \frac{3}{4} & 0 & \frac{1}{4} & \frac{9}{4} \\
    \end{array} \right) \left( \begin{array}{cccc}
      \lambda_1^j & 0 & 0 & 0 \\
      0 & \lambda_2^j & 0 & 0 \\
      0 & 0 & \lambda_3^j & 0 \\
      0 & 0 & 0 & \lambda_4^j \\
    \end{array} \right) \left( \begin{array}{cccc}
      \frac{5}{2} & 0 & \frac{3}{2} & \frac{3}{2} \\
      0 & 2 & 0 & 0 \\
      \frac{3}{4} & 0 & \frac{9}{4} & \frac{1}{4} \\
      \frac{3}{4} & 0 & \frac{1}{4} & \frac{9}{4} \\
    \end{array} \right) \nonumber\\
    &  &  \times \left( \begin{array}{cccc}
      \lambda_1^{N-j} & 0 & 0 & 0 \\
      0 & \lambda_2^{N-j} & 0 & 0 \\
      0 & 0 & \lambda_3^{N-j} & 0 \\
      0 & 0 & 0 & \lambda_4^{N-j} \\
    \end{array} \right) \Biggm] \nonumber\\
    & = & \frac{81}{16}+\frac{9}{8}\left(\frac{\lambda_1}
    {\lambda_4}\right)^j \label{sigmatwor}\,.
\end{eqnarray}
Here, $\lambda_{4}$ and $\lambda_{1}$ are the largest and the
second largest eigenvalues, respectively. The evaluation
was performed in the thermodynamic limit.

The average of the agent at site $i$, $\langle\sigma_i\rangle$, is
evaluated in the same manner:
\begin{equation}
\langle\sigma_i\rangle=\frac{1}{Z}{\rm tr}[AW^N]=\frac{9}{4} \,.
\end{equation}
Then,
\begin{equation}
G(i,i+j)=\frac{9}{8}\left(\frac{\lambda_1}
    {\lambda_4}\right)^j \approx
    e^{-j\ln\left(\lambda_4/\lambda_1 \right)}
    \equiv e^{-j/\xi}\,.
\end{equation}
The correlation function has the same form as in the Ising and
Potts models \cite{yg4,goldenfeld}, with the correlation length
\begin{equation}
\xi=\frac{1}{\ln \left(\lambda_4/\lambda_1 \right)}
 \,.
\end{equation}
Since $\lambda_1=-1+e^{4J/k_BT}$ and
$\lambda_4=1+2e^{J/k_BT}+e^{4J/k_BT}$, the correlation length
becomes
\begin{equation}
\xi=\frac{1}{\ln \left[\left(1+2e^{J/k_BT}+e^{4J/k_BT} \right)/
\left (-1+e^{4J/k_BT} \right) \right] \label{sigmaone}}.
\end{equation}
As in Ising and Potts models, for $H=0$ the two largest
eigenvalues become degenerate at $T=0$, which leads to a divergence
of the correlation length and, therefore, to a zero-temperature
phase transition.

In social terms, the correlation length measures the distance
at which there are relations between agents beyond their own
mean values. This is a causal or influence relationship in the
sense that changing the opinions in one place generates an
influence that causes change up to the correlation length. This
influence operates through local interactions. In terms of social influence the existence of a
correlation length invokes a limit to the propagation of
influence. A return force is a cost for producing fluctuations and this
cost avoids the propagation of
fluctuations beyond a certain distance. When no return force is
present (critical point) fluctuations diverge and influence runs over the whole
society at all scales.

\subsection{Critical exponents}
To get the critical behavior of the thermodynamic properties
one needs to evaluate them near the transition temperature
$T_c$. For 1D models, including the present one, the
transition occurs at $T_c=0$ with exponential singularities
\cite{pathria}. In this case, the usual reduced temperature $t=(T-T_c)/T_c$ is inappropriate.
A different critical point approach
parameter $t=e^{-\Delta/k_BT}$ \cite{pathria} is required to convert the exponential singularities in $T$
into power-law singularities in $t$. The constant $\Delta$ has so far been taken arbitrarily.

Here, we propose that $\Delta$ is given by half the energy
difference between the ground state and the first excited state
of the system. In this way, $\Delta$ eliminates from the value of the exponent any
nonuniversal features, which will be present otherwise.
This convention correctly unifies, independently of the
interaction strength, the Ising and Potts exponents. For the 1D
Ising model the energy difference between the ground state
(say, all spins aligned up) and the first excited state (one
spin aligned opposite to the others) is $4J$; then, for this
case $\Delta=2J$. This agrees with the choice $p=2$ in
$t=e^{-pJ/k_BT}$ \cite{pathria} to bring together the exponents of the 1D
discrete-symmetry models.

For the Potts and thermodynamic Axelrod models, the energy of
the ground state is $-JNF^2$, whereas the energy of the first excited state is
\begin{equation*}
-J \left [(N-2)F^2 + 2(F-1)^2 \right] \,.
\end{equation*}
\noindent Then,
\begin{equation}
\Delta = J(2F-1) \,\,. \label{expt}
\end{equation}
In the case of the Potts model, $F=1$ and $\Delta=J$. This
value of $\Delta$ yields critical exponents of the Potts model
that agree with those of the Ising model \cite{yg4}. For the
thermodynamic Axelrod model, $F=2$ and $\Delta=3J$.

We now can estimate the critical exponents of our
model for the case $F=2$ and $q=2$. We define $h=\mu H/k_BT$
and have $t=e^{-3J/k_BT}$. For $H=0$ and $t\rightarrow0$ the
singular part of the free energy, Eq.~(\ref{freeenergy}),
for a zero-temperature transition \cite{pathria} becomes
\begin{equation}
f(t)=\frac{\mathcal{F}+4NJ}{Nk_BT} \sim t  \,.
\end{equation}
Since $f\sim t^{2-\alpha}$, $\alpha=1$. In the same limit, the
magnetization, Eq.~(\ref{magn}), is
\begin{equation}
m(t)=\frac{M}{N\mu} \sim 1 \,.
\end{equation}
This means from $m \sim t^\beta$ that $\beta=0$. Now, for $t=0$
and $H\rightarrow 0$, the magnetization becomes
\begin{equation}
m(h)=\frac{M}{N\mu} \sim 1 \,.
\end{equation}
Since $m\sim h^{1/\delta}$, the exponent $\delta \rightarrow
\infty$. The low-field susceptibility is obtained from
Eq.~(\ref{suscep})
\begin{equation}
\chi_0(t)=\frac{\chi k_BT}{N\mu^2} \sim \frac{1}{2} t^{-1} \,.
\end{equation}
The susceptibility should go as $t^{-\gamma}$, then $\gamma=1$.
The specific heat, Eq.~(\ref{spheat}), for $H=0$ and
$t\rightarrow 0$ is
\begin{equation}
c(t)=\frac{Ck_BT^2}{NJ^2} \sim -\frac{4}{3}t \,.
\end{equation}
Then, from $C \sim t^\alpha$ one gets $\alpha=1$. Finally, from
Eq.~(\ref{sigmaone}) the correlation lenght
\begin{equation}
\xi(t) \sim \frac{1}{2} t^{-1} \,.
\end{equation}
The correlation length goes as $t^{-\nu}$; then, $\nu=1$.

In conclusion, the differences in the internal space
dimensionality ($F=2$) and interaction strengths of our model
do not alter the Ising universality class. We performed the same calculation
for higher values of $F$ and $q$, obtaining the same results. All the results presented in this section indicate that the 1D thermodynamic Axelrod model exhibits a phase transition at $T=0$.

\section{Thermodynamic and social Axelrod models: a comparison}

\begin{figure}
\scalebox{0.5}{\includegraphics{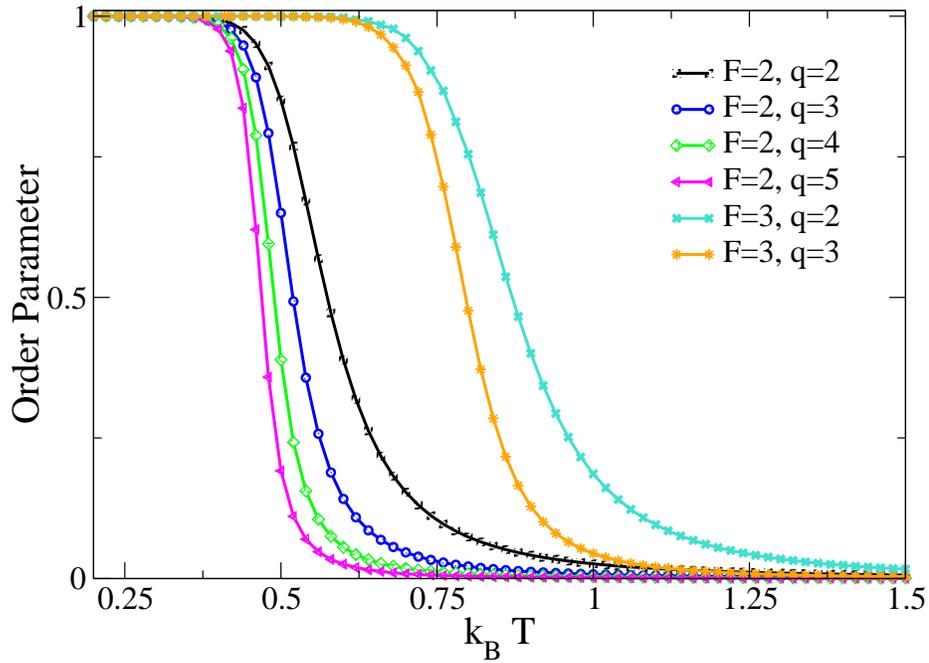}}
\caption{Order parameter
of the thermodynamic Axelrod model for $H=0.005$. The
group of curves to the left correspond to $F=2$ and the ones to
the right to $F=3$. No anomaly is observed as $q$ varies from 2
to 5 for the $F=2$ case. In the nonequilibrium Axelrod model a
crossover from an ordered to a disordered state is seen below
$q=5$ in the 1D case \cite{klemm1}. }
\label{variq}
\end{figure}

We first compare our results with those obtained for the original Axelrod model without the inclusion of any effect. In this case, a monocultural-multicultural (order-disorder) transition is observed at a threshold value $q_c$. Figure \ref{variq} depicts the analytical results for the temperature
dependence of the order parameter
for $F=2$ and $q=2,3,4,5$. For comparison,
we also show curves for $F=3$ and $q=2,3$. A monotonic and
smooth dependence on $q$ is observed for $F=2$, indicating no
phase transition at finite temperature. For both values
of $F$ the order parameter behaviors are similar to those expected
in the Potts model \cite{yg4}. We note that for the analyzed
values of $F$, as $q$ increases the transition becomes gradually
sharper. The data of Fig.~\ref{variq} were obtained, just to be able to perform the numerical calculations, at the very small field of 0.005 $k_B T/ \mu$.

The Axelrod model has also been studied with the addition of noise (cultural drift) \cite{economia,klemm2,toral}. By analyzing the equilibrium configurations and their stability, it was found that below a critical value of the noise rate $r_c$ (noise is introduced as a perturbation of a single agent at certain times during the dynamics) the initially multicultural population converges to a monocultural one, whereas for $r>r_c$ the system moves to a multicultural state  \cite{economia}. The threshold value $r_c$ depends on the system size and is almost independent of $q$. In 1D $r_c\sim 1/N^2$ \cite{economia,toral}, thus in the limit $N\rightarrow \infty$, the value $r_c\rightarrow 0$, implying that in this condition there is no phase transition and that for any finite $r$ the system converges to a multicultural state \cite{economia,toral}. This behavior is analog to the one found in the thermodynamic Axelrod model, in which for any finite $T$ the system goes to a multicultural state. The polarized state for
any $r>0$ is equivalent to the zero-magnetization phase for any finite $T$ in discrete-symmetry systems such as the present one. The noise variable $r$ corresponds to temperature $T$.

The implications of the thermodynamic limit ($N\rightarrow\infty$) for the noisy Axelrod model are known \cite{economia,klemm2,toral}. Our study, focused on the equilibrium approach, underlines a possible connection between the thermodynamic model and the nonequlibrium noisy Axelrod model. Universal behavior is equivalent once one equates noise and thermal fluctuations. It is worthwhile to study then the present Hamiltonian model in higher dimensions to verify the latter assertion. It is also interesting to see whether the energy tendency to increase similarity of agents can change the universality class from that of Potts.

Regarding the effect of a magnetic field; at $T=0$, a 1D thermodynamic system is always magnetized for any value of the applied field. Even though a direct comparison is only qualitative, it is worth mentioning that in the original Axelrod model an external field (mass media) gives rise to related behavior in a 2D system. By using, as the order parameter, the average fraction of cultural clusters $g=\langle N_g \rangle /N$, where $N_g$ is the number of clusters formed in the final state, it was suggested that, in 2D, mass media can induce a multicultural state when its strength is above a certain threshold \cite{gonzalez1,gonzalez2}. This was also shown with the order parameter $\phi$ \cite{yg}. Using the order parameter $g$, Peres and Fontanari \cite{fontanari} showed that in 2D the effect of mass media on the Axelrod model always (for any value of $H$) displays a tendency toward cultural diversity when $N\rightarrow\infty$. That is to say, no order-disorder transition is observed with $g$ in such limit.  On the other hand, when the normalized size of the largest cluster $\phi$ is used as the order parameter (the one applied in our work) in the original Axelrod model, the results indicate that the effect of mass media, inducing a multicultural phase, persists in the limit $N\rightarrow\infty$ \cite{thesisavella}. Hence, it would be of interest to study the 2D thermodynamic Axelrod model with both order parameters $\phi$ and $g$; in particular with $g$. Of course, the $N\rightarrow\infty$ limit is mostly academic in the social context, since real social systems are small compared to thermodynamic systems.

\section{Nature of thermal society of agents}

Two basic questions come to mind when proposing a thermodynamic
model for society, namely: what is the meaning of temperature
and detailed balance? In the Axelrod model each agent is
described as a vector whose $F$ components can take $q$ values
that reflect the variety of postures an individual can have on
a particular scope of action in society. Temperature may be
thought as related to an energy scale that competes with the
regular interactions between individuals; a high temperature
renders their interactions moot, while a low temperature leads
to a domination of the individual interactions and to the
system settling into a minimum energy state. In this sense, in
a society at high temperatures individuals would have
fluctuating positions with a small relation to those of their
circle of interaction, whereas at low temperatures individuals
would pay much attention to their circle of interaction and
tend to lower posture differences. In terms of a statistical
ensemble approach, a group of individuals of the society would
be subjected to a temperature reservoir that competes with
individual interactions.

What sets this temperature scale might be any motive for
confusion, for speculation or for uncertainty that could
disrupt bonds between individuals, the agreement is not
conducive to the benefit of the individual. This would be consistent
with regarding temperature as a parameter coupled to entropic effects. In social
terms, as the temperature increases the agents communicate less
effectively, their coupling decreases, they are less convincing,
in a sense, to their neighbors who choose or preserve their own
positions without regard to their peers. This has entropic
benefits since there is an increasing amount of ways people can
disagree.

Thermal equilibrium is a situation where some Helmholtz-type
free energy is minimal, and energy fluctuations exist subject
to the condition of detailed balance. From Eq.~(\ref{dbe}) one
can see that detailed balance dictates that an individual in a
highly probable state $n$ should balance with an individual in
a less probable state $m$. It is reasonable to expect that it
is easier for the individual in state $m$ to adjust itself to
the mainstream rather than the other way around. This can be
seen as peer pressure or pressure to conform to the norm. This
is an interesting perspective in the sense that varying
temperature can yield thresholds for phase changes and set
tipping points for collective behavior. On the other hand, a
single temperature may not be set for different scopes of
action, since a religious posture is certainly less fluctuating
than a political posture. In our model this may be taken into
account by weighting the interaction factor, $J_{ij}$, that depends
on the cultural-featured $F$ vector.

\section{Summary}

We have presented a thermodynamic counterpart of the Axelrod model
of social influence. The transfer matrix method was used to exactly
solve for the thermodynamic and critical properties of the 1D model. A unifying proposal was made to compare exponents across different
1D discrete models, since current definitions depend on
model details. We have also interpreted the implication of thermodynamic cultural
dissemination.

The differences in internal symmetry and interaction in the thermodynamic Axelrod model
with respect to the Potts model where not relevant from the point of view of criticality.
An order-disorder phase transition occurs at $T=0$ independent of the cultural trait
$q$ and feature $F$ variables. Our model, emulating Axelrod rules, increases the tendency to share
cultural traits as the agents are more similar. This feature precipitates ordering in the system
toward lower energy (T=0) as compared to Ising-like Hamiltonians. We expected that this feature would bring about absorbing-state type behavior, but in ring-like topologies fluctuations would be
too strong so there would be only multicultural states for all finite $T$. Although the new Hamiltonian has a state-dependent bonding interaction, the model belongs to the Ising and Potts universality class.

The comparison with a Hamiltonian system sheds some light on the effects of two of the main ingredients of the out-of-equilibrium Axelrod model in the 1D case. The presence of mass media carries the system only to nonuniform regime. On the other hand, the thermodynamic
counterpart also revealed that the original out-of-equilibrium 1-D Axelrod model in the limit $N\rightarrow\infty$ remains in a disordered state for any finite noise, as expected for any 1D interacting particle system. It would be interesting to study the model in higher dimensions, where energy related effects can have an increasingly stronger role as compared to entropic features.

\section{Acknowledgments}
Y.G. thanks support from the Venezuelan Government's project Misi\'{o}n Ciencia and the Instituto Venezolano de Investigaciones Cient\'{i}ficas (IVIC). I.B.
appreciates the financial assistance from IVIC through project
No. 441. Y.G. is thankful for assistance from the Condensed Matter
Laboratory at Universidad Sim\'{o}n Bol\'{i}var.


\section{References}

\begin{thebibliography}{14}
\bibitem{axelrod} R. Axelrod, J. Conflict Resolut. 41, 203 (1997).
\bibitem{castellano2} C. Castellano, S. Fortunato, V. Loreto, Rev. Mod. Phys. 81, 591 (2009).
\bibitem{barrat} A. Barrat, M. Barthélemy, A. Vespignani, \emph{Dynamical Processes on Complex networks} (Cambridge University Press, Cambridge, 2008).
\bibitem{castellano1} C. Castellano, M. Marsili, A. Vespignani, Phys. Rev. Lett. 85, 3536 (2000).
\bibitem{gonzalez1} J.C. Gonzalez-Avella, M.G. Cosenza, K. Tucci, Phys. Rev. E 72, 065102 (2005).
\bibitem{gonzalez2} J.C. Gonzalez-Avella, V.M. Eguiluz, M.G. Cosenza, K.Klemm, J.L. Herrera,
M. San Miguel, Phys. Rev. E 73, 046119 (2006).
\bibitem{yg} Y. Gandica, A. Charmell, J. Villegas-Febres, I. Bonalde, Phys. Rev. E 84, 046109 (2011).
\bibitem{economia} K. Klemm, V. M. Egu\'{i}luz, R. Toral, M. San Miguel, J. Econ. Dyn. Control 29, 321 (2005).
\bibitem{klemm2} K. Klemm, V. M. Egu\'{i}luz, R. Toral, M. San Miguel, Phys. Rev. E 67, 045101 (2003).
\bibitem{toral} R. Toral, C. J. Tessoni, Commun. Comput. Phys. 2, 177 (2007).
\bibitem{hammal} O. Al Hammal, H. Chat\'{e}, I. Dornic, M. A. Mu\~{n}oz, Phys. Rev. Lett. 94, 230601 (2005).
\bibitem{report} B. J. West, E. Geneston, P. Grigolini, Phys. Rep. 468, 1-99 (2008).
\bibitem{Nonequilibrium} M. Henkel, M. Pleimling, \emph{Non equilibrium phase transitions
Volume 2: Ageing and Dynamical Scaling far from equilibrium} (Springer, Heidelberg, 2010).
\bibitem{diu} B. Diu, C. Guthmann, D. Lederer, B. Roulet, \emph{Physique Statistique} (Hermann, Paris, 1989).
\bibitem{kardar} M. Kardar, \emph{Statistical Physics of Fields} (Cambridge University Press, Cambridge, 2007).
\bibitem{pathria} R. K. Pathria, \emph{Statistical Mechanics} (Butterworth-Heinemann, Oxford, 1996), 2nd ed.
\bibitem{goldenfeld} N. Goldenfeld, \emph{Lectures on Phase Transitions and the Renormalization Group} (Perseus Books, Reading, 1992).
\bibitem{yg4} Y. Gandica, Ph.D. thesis, Instituto Venezolano de Investigaciones Cient\'{i}ficas (2012).
\bibitem{klemm1} K. Klemm, V. M. Egu\'{i}luz, R. Toral, M. San Miguel, Physica A 327, 1 (2003).
\bibitem{fontanari} L. R. Peres, J. F. J. Fontanari, J. Phys. A: Math. Theor. 43, 055003 (2010).
\bibitem{thesisavella} J. C. Gonzalez-Avella, Ph.D. thesis, Universitat de les Illes Balears (2010).

\end{thebibliography}
\end{document}